\documentclass[conference]{IEEEtran}

\usepackage{amsmath,amssymb,epsfig,cite,footnote,authblk}

\newcommand{\pr}[1]{\left( #1\right)}
\newcommand{\prr}[1]{\left[ #1 \right]}
\newcommand{\es}[1]{\begin{equation}\begin{split}#1\end{split}\end{equation}}

\newcommand{\R}{\mathbb{R}}
\newcommand{\N}{\mathbb{N}}

\newcommand{\V}{\mathcal{V}}

\newcommand{\erfc}{\mathrm{erfc}}

\newcommand{\bt}{\mathbf{t}}
\newcommand{\dd}{\textrm{d}}

\begin{document}

\title{Directional antennas improve the link-connectivity of interference limited ad hoc networks}
\author[1]{Orestis Georgiou}
\author[1]{Shanshan Wang}
\author[1]{Mohammud Z. Bocus}
\author[2]{Carl P. Dettmann}
\author[3]{Justin P. Coon}
\affil[1]{Toshiba Telecommunications Research Laboratory, 32 Queens Square, Bristol, BS1 4ND, UK}
\affil[2]{School of Mathematics, University of Bristol, University Walk, Bristol, BS8 1TW, UK}
\affil[3]{Department of Engineering Science, University of Oxford, Parks Road, OX1 3PJ, Oxford, UK}
\maketitle

%\pagestyle{plain}
%%%%%%%%%%%%%%%%%%%%%%
%\thispagestyle{fancy}

\begin{abstract}
We study wireless ad hoc networks in the absence of any channel contention or transmit power control and ask how antenna directivity affects network connectivity in the interference limited regime.
We answer this question by deriving closed-form expressions for the outage probability, capacity and mean node degree of the network using tools from stochastic geometry.  
These novel results provide valuable insights for the design of future ad hoc networks. 
Significantly, our results suggest that the more directional the interfering transmitters are, the less detrimental are the effects of interference to individual links.  
%This observation assumes that all nodes are randomly positioned in $\R^2$ with antennas randomly oriented in $[0,2\pi]$.
We validate our analytical results through computer simulations.
\end{abstract}

%%%%%%%%%%%%%%%%%%%%%
\section{Introduction \label{sec:intro}}
With over a decade of research and technological advancements, ad hoc networking has evolved into a mature and promising technique for future communication systems \cite{frodigh2000wireless}.  Some of the advantages of ad hoc networks include increased coverage through multihop transmission, improved capacity and the ability to autonomously organise and initiate communication.  On that account, such systems have found applications in sensor networks and are very attractive for rapid deployment in disaster scenarios \cite{cordeiro2011ad}.   

While the benefits are numerous, the inherent structure of ad hoc networks leads to an interference limited regime \cite{dousse2005impact}.
One of the main contributors for such a regime is the widespread use omni-directional antennas in the network.  Traditionally, omni-directional antennas have been favoured for wireless ad hoc networks due to the cost and size limitations.  More recently, directional antennas have been considered to improve the spectral reuse and efficiency as well as control the level of interference in the systems, see e.g., \cite{Ramanathan2005, Bazan2012, Doufexi2004, Janecek2009, Fawaz2008} and references therein.  A directional antenna system usually comprises of an adaptive antenna array and control algorithms to control the overall radiation pattern of the transceiver.  Such characteristics and their potentials have been extensively studied in the literature.  In \cite{Ramanathan2005} for instance, a complete method for ad hoc networking with directional antennas is presented.  The authors show that directional antennas can provide a significant improvement in network capacity.  A concise overview on medium access control (MAC) protocols for wireless ad hoc networks with beamforming antennas is presented in \cite{Bazan2012}.  The work presented in \cite{Kubichek2014} on the other hand investigates the array design for mobile ad hoc networks that can improve the performance of the networks.  Along similar lines, the use of sectorised antennas in ad hoc networks have been investigated in \cite{Doufexi2004} and \cite{Janecek2009}.  

A radically different approach from the above was considered in \cite{bettstetter2005does}
and was further generalised and analysed in \cite{coonanisotrop, georgiou2013connectivity, georgiou2015multihop}. Here, all transmitting directional antennas are assumed to be oriented in a random direction and do not attempt any adaptation or control; a reasonable assumption in ad hoc networks with hardware-limited and power-constrained wireless devices.
The above studies and many others \cite{vilzmann2005hop, zhou2007analytical, zhang2012analysis,dai2015delay} on randomly oriented directional antennas or random beamforming schemes (where nodes  beamform in a random direction) suggest that simple schemes can lead to network performance gains in routing, end-to-end delay, reachability, and capacity.
Implicit for this return, is perfect interference management and a good MAC.

The contribution of this paper differs in this regard as we assume that nodes can access the channel without contending for it first and therefore interference from concurrent transmissions can be a dominant factor.
The ALOHA access protocol for example can be modelled by assuming that each node transmits randomly, irrespective of any nearby transmitter.
Such a random access scenario \cite{joo2009performance} is particularly relevant in disaster scenario for instance where deployed nodes may not be aware of the existence of other infrastructures.  In these and similar cases \cite{baccelli2009stochastic}, interference can be a limiting factor, especially in the dense regime.  We demonstrate and quantify using tools from stochastic geometry \cite{haenggi2009stochastic} that ad hoc networks equipped with directional antennas have a higher tolerance to interference and thus can operate reliably at higher densities and attain higher throughputs.

We first study the outage probability and connectivity of an interference limited Poisson network where transceivers are equipped with directional antennas.  
Closed-form expressions for the per-link outage probability (c.f., \eqref{ClosedFormHij} herein) are derived highlighting how this metric depends on antenna directivity, path loss exponent, and interference caused by concurrent transmissions.  The analysis indicates that the interference caused by transmitters with directional antennas is typically less detrimental to point-to-point transmissions as compared to the case of isotropic antennas.
We further present analytic expressions in closed-form for the mean degree of interference limited ad hoc networks.  The results show that directivity reduces the number of nodes a transmitter can simultaneously communicate to in a broadcast fashion. This trade-off between link quality and number of links is intuitively expected.  However, to the best of the authors' knowledge, no quantitative analysis have been conducted on that matter before.  Using the derivations presented in this contribution, communication parameters can be tuned more appropriately to meet the design criteria and are thus of interest to both network theoreticians and wireless practitioners.

The rest of the paper is organised as follows.  The system model and definitions are introduced in Section \ref{sec:model}.  The performance analysis of the system is presented in Section \ref{sec:Analysis}.  Numerical results are provided in Section \ref{sec:num} and Section \ref{sec:con} provides some concluding remarks.

%%%%%%%%%%%%%%%%%%%%%
\section{Network Definitions and System Model \label{sec:model}}

\subsection{Poisson Network Deployment}
We consider an ad hoc network deployed in the $\R^2-$domain where all wireless devices are uniformly and randomly distributed according to a two-dimensional Poisson point process (PPP) with intensity $\rho$.  We further assume that no carrier sensing is employed for medium access.  In any finite subset $\V\subset\R^2$ of area $V=|\V|$, there are a total of $N$ transmitters where $N$ is a Poisson distributed random variable with mean $\rho V$.  We also assume that there is no power control and all devices transmit with the same power $\mathcal{P}$.  For the purpose of the analysis and due to the homogeneity of the PPP, we consider that the receiving node is located at the origin of our coordinate system.  The locations of the transmitting nodes are denoted by the vector $\bt_i \in \R^2$ in polar coordinates for $i=1,2,\ldots$, while the distance between a transmitter and the receiver at the origin is given by $t_i=|\bt_i|$.
We also assume that the transmitting nodes are active in every resource block.

\subsection{Path-loss and Fading}
The signal-to-noise ratio (SNR) is a commonly used metric to quantify the quality of a communication link.  This value is dependent on path loss attenuation. Since the signal power received by a destination node in the far field is inversely proportional to the separation distance, a simple path loss attenuation function $g(x)$ is often adopted to aid mathematical analysis 
\es{
g(x)= \frac{1}{x^\eta + \epsilon} , \qquad 2<\eta,\, \quad 0 \leq \epsilon 
\label{pathloss}
,}
where $\eta$ is the path loss exponent and is usually greater than $2$ for non-line-of-sight urban environments.  Notably, it marks the signal attenuation between a transmitter and receiver. The $\epsilon$ term included in the denominator above is applied to ensure that the attenuation function non-singular at zero distance. 

In addition to the path loss attenuation, small-scale fading also affects the received signal power. Rayleigh fading is a typical small scale fading model adopted by the scientific community which is simple enough to maintain mathematical tractability whilst producing insightful engineering results.
We adopt this approach here and model the channel gain between transmitting node $i$ and the receiver by $|h_{i}|^2$.  This term is an exponentially distributed random variable of mean one.
%We will not consider the effect of large-scale fading (e.g. shadowing) in this paper but intend to do so in a dedicated publication in the future. 

\subsection{Antenna Directivity}
\begin{figure}[t]
\centering
\includegraphics[scale=0.3]{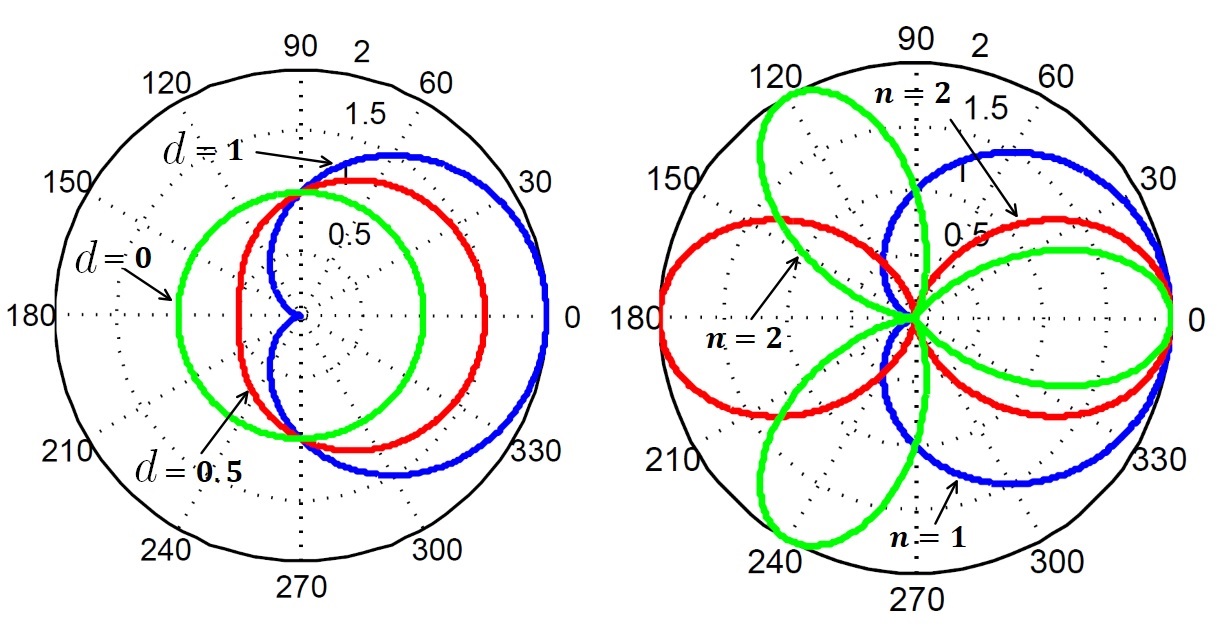}
\caption{
\textit{Left:} Three antenna radiation patterns for different directionality factors $d $ as given by \eqref{antennagain} ($n=1$) . 
\textit{Right:} Three antenna radiation patterns for different number of lobes $n$ as given by \eqref{antennagain} ($d =1$). 
}
\label{fig:direcantenna}
\end{figure}
To achieve a more realistic model, the isotropic antenna model commonly assumed is substituted here by that of directional antennas with variable beam-width. Indeed most of the scientific literature concerned with network connectivity either use unrealistic, perfectly isotropic antennas or adopt crude sectorised radiation patterns.
For mathematical tractability, we will ignore small side-lobes and back-lobes and use the following smooth function to model the gain of a multi-directional antenna \cite{balanis2012antenna}
\es{
G(\theta) =1 + d  \cos( n \theta), \qquad 0 \leq d  \leq 1 , \quad n\in\N^+
\label{antennagain}
.}
The gain function $G$ represents the ratio between the signal intensity in direction $\theta$, and the signal intensity had the same power been radiated or received using an isotropic antenna. 
We use $G$ to represent the transmitting antenna gain and $\bar{G}$ for the receiving antenna gain.
The variable $n=1,2,\ldots$ controls the number of main lobes that the antenna can achieve while $d $ regulates the beamwidth of the main lobes. 
For $d =0$, an isotropic radiation pattern is obtained (i.e., the antenna gain is equal in all directions) while for $d >0$ the antenna has $n$ preferred radiation directions (see Fig.~\ref{fig:direcantenna}).  The reader should note that $\int_0^{2\pi} G(\theta) \dd \theta =2\pi$. 

\subsection{Interference and SINR}
In Poissonian networks, wireless devices transmit with no channel access coordination such that signals interfere at the receiver.  This can lead to packet losses or lower data rates. Such occurrences can be considered as the worst-case scenario where no interference management (e.g., collision avoidance) techniques are employed.
It is therefore important to consider the signal-to-interference-plus-noise ratio (SINR) of transmitter $i$ as received by the receiver located at the origin
\es{
\textrm{SINR}_i = \frac{\mathcal{P} |h_{i}|^2 g(t_{i}) G_{i} \bar{G}_{i}}{\mathcal{N} +  \gamma\mathcal{I}}
\label{SINR}
,}
where $\mathcal{N}$ is the long time average background noise power and $\mathcal{I}$ is the total interference received at the origin which we treat as a shot-noise, i.e., the sum of all unwanted signals from nodes which are transmitting at the same time or using the same spectrum resource as transmitting node $i$.  This value is computed as 
\es{
\mathcal{I} = \mathcal{P} \sum_{k\not=i}  |h_{k}|^2 g(t_{k}) G_{k} \bar{G}_{k}
\label{Interference}
.}
The orthogonality factor $\gamma\in[0,1]$ denotes the proportion by which a desired signal is influenced by interference. 
This can be interpreted as the spreading gain in CDMA systems whose value depends on the spreading codes used. $\gamma=0$ represents an interference-free system where $\textrm{SINR}_{i} = \textrm{SNR}_{i}$. On the opposite extreme, $\gamma =1$ denotes an interference-rich environment where all communicating nodes interfere with one another.

\begin{figure}[t]
\centering
\includegraphics[scale=0.22]{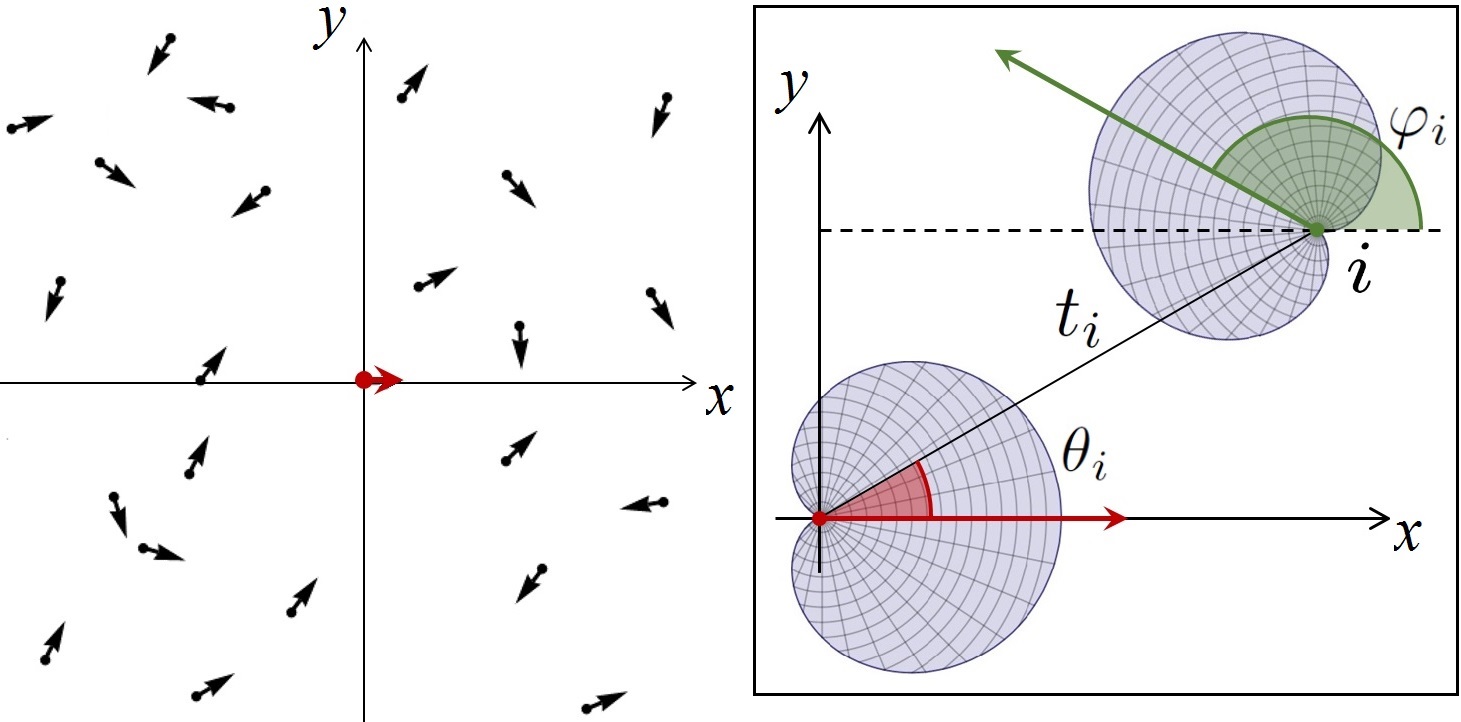}
\caption{
\textit{Left:} Schematic of the setup showing a receiver at the origin oriented towards the $x$-axis, and 23 randomly located and randomly oriented transmitters.
\textit{Right:} Schematic of the coordinate system used relative to the receiver at the origin. 
Transmitter $i$ is located at $\bt_i= (t_i, \theta_i)$ using polar coordinates and its antenna is oriented towards $\varphi_i$, such that the gain angles are $\theta_i$ for the receiver's gain, and $\theta_i+\pi-\varphi_i$ for the transmitter's gain.
}
\label{fig:setup}
\end{figure}

\subsection{Randomly oriented directional antennas}
Due to the broadcast nature of wireless communications, transmissions can always be interfered with by other on-going transmissions.  
It is intuitive that the SNR between a perfectly aligned directional transmitter-receiver node pair is significantly improved compared to case with isotropic antennas.  How such a communication link is affected by concurrent interfering transmissions is however not a well explored area.
We therefore study the SINR of a transmitter-receiver node pair which is perfectly and imperfectly aligned in the presence of a large number of randomly oriented directional interferers. 
This set-up is shown in Fig.~\ref{fig:setup} and is motivated by the randomness or unpredictability of the interfering antenna orientations in a dense urban (possibly mobile) environment.  Such conditions may be found in randomly deployed/dropped wireless sensor networks as described in \cite{cordeiro2011ad}.

%%%%%%%%%%%%%%%%%%%%%%%%%%%%%%%%%%%%%%%%%
\section{Connectivity Analysis \label{sec:Analysis}}

\subsection{Connection Probability\label{subsec:hij}}
The connection probability is the complement of the outage probability and is a fundamental quality metric for point-to-point links. 
It can be derived from \eqref{SINR} by setting a minimum reception threshold $q$ on $\textrm{SINR}_{i}$ for the receiver to successfully decode the signal from transmitter $i$. 
We define the connection probability by
\es{
H_{i}=\mathbb{P}[\mathrm{SINR}_{i} \geq q]
= \mathbb{P}\Big[|h_{i}|^2 \geq \frac{q (\mathcal{N}+  \gamma\mathcal{I})}{\mathcal{P} g(t_i) G_{i} \bar{G}_{i}}\Big]
\label{hij}.}
The connection probability can be interpreted as the probability that the received $\textrm{SINR}_{i}$ exceeds the threshold $q$, or the probability that this connection can be set up successfully. 
%All possible transmitter and receiver pairs follow Poisson Point Processes (PPPs) on $\R^2$.

Conditioned on the interference term $\mathcal{I}$ and using the fact that $|h_{i}|^2\sim\textrm{exp}(1)$, we can write \eqref{hij} as
\es{
H_{i} &= \mathbb{E}_{\mathcal{I}} \Big[ \mathbb{P} \prr{|h_{i}|^2\geq \frac{q(\mathcal{N}+\gamma\mathcal{I})}{\mathcal{P}g(t_i)G_{i} \bar{G}_{i}}} \, \Big| \, \mathcal{I} \Big] = \mathbb{E}_{\mathcal{I}}\! \prr{  e^{ -\frac{q(\mathcal{N}+\gamma\mathcal{I})}{\mathcal{P}g(t_i)G_{i} \bar{G}_{i}}}}\\
&= e^{-\frac{q \mathcal{N}}{\mathcal{P}g(t_i)G_{i} \bar{G}_{i}}}\mathcal{L}_{\mathcal{I}} \pr{ \frac{s}{\mathcal{P}}}
\label{Laplace}
,}
where $s= \frac{q\gamma}{g(t_i)G_{i} \bar{G}_{i}}$ and
\es{
\mathcal{L}_{\mathcal{I}}\pr{ \frac{s}{\mathcal{P}}}\! &= \! \mathbb{E}_{\mathcal{I}}\prr{e^{-\frac{s\mathcal{I}}{\mathcal{P}}} } \! = \! \mathbb{E}_{|h_{k}|^2, t_k, G_k}\!\! \prr{e^{-s \sum_{k\not= i}|h_{k}|^2 g(t_k) G_{k} \bar{G}_{k}}}\\
%&= \mathbb{E}_{|h_{kj}|^2,d_{kj},\varphi _k}\prr{ \prod_{k\not=i}^{N_T-1} e^{-s |h_{kj}|^2 g(d_{kj})G_{kj} G_{jk}}  } \\
&= \mathbb{E}_{t_k,G_k} \Big[ \prod_{k\not=i} \mathbb{E}_{|h_{k}|^2} \Big[ e^{-s |h_{k}|^2 g(t_k)G_{k} \bar{G}_{k}} \Big]  \Big]
\label{LaplaceTrans}
}
is the Laplace transform of random variable $\mathcal{I}$  evaluated at $\frac{s}{\mathcal{P}}$.
In the last line of \eqref{LaplaceTrans} we have used the fact that the locations of the interferers $t_k$, orientations $G_k$ and the channel gains $|h_k|^2$ are all independent random variables.

The probability generating function of a  homogeneous PPP $\Xi$ in some space $\mathcal{S}$ with intensity function $\lambda$ satisfies  
\es{
\mathbb{E} \Big[ \prod_{\xi \in \Xi} f(\xi) \Big] = \exp\pr{ - \lambda \int_{\mathcal{S}} \pr{ 1-f(\xi) }  \dd \xi  }
\label{PGF},}
for functions $f$ under the circumstance that $0<f(\xi)<1$.
To progress in the derivations, we combine the two dimensional spatial PPP of transmitter locations with their random antenna orientations to form a three dimensional homogeneous PPP in $\mathcal{S}= \R^2 \times [0,2\pi]$ of intensity $\lambda= \frac{\rho}{2\pi}$ and use \eqref{PGF} to arrive at
\es{
\mathcal{L}_{\mathcal{I}} \pr{\frac{s}{\mathcal{P}}}\! &=\! \exp \Big( - \frac{\rho}{2\pi} \int_0^{2\pi}\!\!\int_{\R^2} \!\Big(1- f(\bt_k)\Big) \dd \bt_k \dd \varphi_k \Big) \Big]
\label{Sigma1}
}
where the outer integral averages over all possible orientations of transmitting antenna $k$ given by its angular difference from the $x$-axis given by $\varphi_k$.
In \eqref{Sigma1} we have also defined 
\begin{equation}
f(\bt_k)=\mathbb{E}_{|h_{kj}|^2} \Big[ e^{- \zeta |h_{kj}|^2 } \Big]\! =\! \int_{0}^{\infty} \!\! e^{- z(\zeta +1)} \dd z = \! \frac{1}{1+ \zeta},
\end{equation}
and $\zeta=s g(t_k) G_{k} \bar{G}_{k}$. 
The integrals in \eqref{Sigma1} can be expanded into
\es{
I&=\!\int_{0}^{2\pi}\!\! \int_0^{2\pi}\!\! \int_0^\infty \!\!\frac{ s g(t_k) G(\theta_k+\pi - \varphi_k) \bar{G}(\theta_k) t_k}{1+ s g(t_k) G(\theta_k+\pi - \varphi_k) \bar{G}(\theta_k) }  \dd t_k \dd \theta_k \dd \varphi_k \\
&= \!\int_{0}^{2\pi}\!\! \int_0^{2\pi} \frac{\pi \sigma (\epsilon +\sigma)^{\frac{2}{\eta}-1}  }{\eta \sin \frac{2\pi}{\eta}}  \dd \theta_k  \dd \varphi_k 
\label{Sigma3}
,}
where $\sigma = s G(\theta_k+\pi - \varphi_k) \bar{G}(\theta_k)$.
Notice that the receiver's gain $\bar{G}$ depends only on the angular position $\theta_k$ of transmitter $k$, whilst the transmitter's gain $G$ depends on its antenna orientation $\varphi_k$ relative to its angular position with respect to the receiver at the origin. 

\begin{figure}[t]
\centering
\includegraphics[scale=0.3]{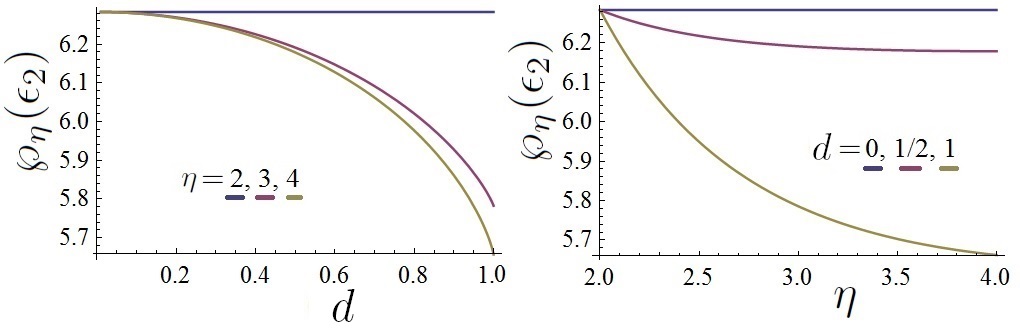}
\caption{
Plots of $\wp_\eta(d )$ given in \eqref{ClosedFormGain} as a function of the directionality factor $d $ (left) and the pathloss exponent $\eta$ (right) for different values  as indicated in the figures.
}
\label{fig:wp}
\end{figure}

Setting $\epsilon=0$ we may further simplify the above integrals to obtain
\es{
I= \frac{\pi s ^{\frac{2}{\eta}}}{\eta \sin\frac{2\pi}{\eta}} \Big( \int_{0}^{2\pi} G(\theta)^{\frac{2}{\eta}} \dd \theta \Big)^2\\
\label{ClosedFormInter}
}
where we have used the $2\pi$ periodicity of the gain function $G$ and require that $\eta>2$.
The integral in \eqref{ClosedFormInter} can further be simplified to
\es{
\wp_\eta(d )\!= \!\int_{0}^{2\pi} \!\!\! G(\theta)^{\frac{2}{\eta}} \dd \theta 
&=\pi \Big[ (1-d )^{\frac{2}{\eta}} {}_2 F_1 \Big( \frac{1}{2} , -\frac{2}{\eta} , 1 , \frac{2d }{d  -1} \Big) \\
&+ (1+d )^{\frac{2}{\eta}} {}_2 F_1 \Big( \frac{1}{2} , -\frac{2}{\eta} , 1 , \frac{2d }{d +1} \Big)\Big] \\
&= 2\pi - \frac{\pi(\eta-2)}{\eta} d ^2 + \mathcal{O}(d ^4)
\label{ClosedFormGain}
,}
where ${}_2 F_1 $ is the Gauss hypergeomeric function.
For the fully anisotropic case of $d=1$ we have
\es{
\wp_\eta(1)= \frac{ 2^{\frac{2}{\eta}} \eta \sqrt{\pi}   \Gamma(\frac{1}{2}+\frac{2}{\eta})}{\Gamma(\frac{2}{\eta})} = 2\pi -  \frac{2\pi \ln 4}{\eta} +\mathcal{O}(\eta^{-2}) 
.}
Notice that there is no $n$ dependence in \eqref{ClosedFormGain} and that $\wp_\eta(d )<2\pi$ is a decreasing function of the antenna directivity $d $ (see Fig.~\ref{fig:wp}).
Back substitution leads to
\es{
H_{i}=\exp \Big(-\frac{q \mathcal{N} t_i^\eta}{\mathcal{P} G_{i} \bar{G}_{i}}\Big) \exp \Big(-\rho   \frac{ t_i^2 \wp_\eta(d )^2}{2\eta \sin\frac{2\pi}{\eta}}\pr{\frac{q \gamma}{G_{i} \bar{G}_{i}}}^{\frac{2}{\eta}}  \Big)
\label{ClosedFormHij}
,}
for the case of $\epsilon=0$.

\textit{Remark 1: Expression \eqref{ClosedFormHij} is the main analytic result of this paper.  This result is validated through computer simulations in Sec.~\ref{sec:num}.
Notice that the connection probability $H_i$ depends on the radiation pattern of the transmitter $G(\theta_i+\pi-\varphi_i)$, the receiver $\bar{G}(\theta_i)$ as well as the relative distance between them denoted by $t_{i}$.
Significantly however notice that $H_i$ depends exponentially on the directivity of the interfering nodes measured indirectly through $\wp_\eta(d )$.
Namely, \textit{ceteris paribus}, the more directional the interfering nodes, the smaller the value of $\wp_\eta(d )$, and hence the less detrimental are the effects of interference to the connection probability $H_i$. 
}
% And on the basis of $H_{ij}$, more expressions for theoretical achievable date rate and spatial density of successful transmissions are obtained in the following sections.

%%%%%%%%%%%%%%%%%%%%
\subsection{Maximum Theoretical Data Rate}
According to the Shannon-Hartley theorem, the capacity of a band limited AWGN channel with bandwidth $B$ is $C=B \log_2(1+\textrm{SNR})$. 
Assuming adaptive modulation and coding is used, the theoretical maximum  data rate in nats per second can be expressed as
\es{
v_{i}=\mathbb{E}[\textrm{ln}(1+\textrm{SINR}_{i})].
\label{drdefinition}
}
The maximum data rate is evaluated upon a vast number of transmissions with fading distributions and spatial distribution of interfering devices. 
It is calculated by $\mathbb{E}[X]=\int\mathbb{P}[X>x] \dd x$, which therefore leads us to
\es{
v_{i}&=\int_{0}^{\infty} \mathbb{P} \Big[ \textrm{ln}\Big( 1+ \frac{\mathcal{P}|h_{i}|^2 g(t_i)}{G_{i} \bar{G}_{i}(\mathcal{N}+\gamma \mathcal{I})}\Big)>x \Big]\dd x\\
&=\int_{0}^{\infty} \exp\Big(-\frac{\hat{q}\mathcal{N}}{\mathcal{P}g(t_i) G_{i} \bar{G}_{i}}\Big) \mathcal{L}_{\mathcal{I}}\Big(\frac{\hat{q}\gamma}{\mathcal{P} g(t_i) G_{i} \bar{G}_{i}}\Big)\dd x
\label{datarate}
}
where we have used the substitution $\hat{q}=e^x-1$ and $\hat{s}=\frac{\hat{q}\gamma}{g(t_i)G_{i}\bar{G}_{i}}$. 
It is therefore clear that the theoretical maximum data rate also relies on the relative position of transmitter and receiver and antenna radiation pattern of the wireless devices in a given environment. 

\textit{Remark 2: Moreover and similarly to $H_i$, the more directional the interfering nodes, the smaller the value of $\wp_\eta(d )$, and hence the less detrimental are the effects of interference to the theoretical data rates $v_i$ for transmissions from $\bt_i$ to the receiver at the origin.}

%%%%%%%%%%%%%%%%%%%%%%
\subsection{Mean Degree of Successful Transmissions}
Another metric of interest for measuring the local connectivity of ad hoc networks is the mean node degree. 
This is defined as the average number of nodes per unit area from which a typical receiving node can successfully decode the transmissions. 
Alternately, the mean degree can be interpreted as the typical number of potential neighbors that one node can connect to in a given time slot.
Note that for symmetric channels and identical transmitter-receiver antennas, the mean in-degree equals the mean out-degree.  Hence this gives us an indication on how many receivers a typical transmitter can successfully connect to per transmission.
Therefore the node degree is hereby defined by
\es{
\mu = \frac{\rho}{2\pi} \int_0^{2\pi}\int_{\R^2}H_{i} \dd \bt_i \dd \varphi_i
\label{NodeDegreeDenfi}
}
Setting $\eta=4$ and $\epsilon=0$, \eqref{NodeDegreeDenfi} can be evaluated in closed-form as
\es{
\mu &= \frac{\rho}{2\pi} \! \int_{0}^{2\pi} \!\! \int_{0}^{2\pi} \!\! \int_{0}^{\infty}\!\! H_{i} t _i \dd t _i \dd \theta_i \dd \varphi_i =\frac{2}{\sqrt{\pi q \gamma}} z e^{z^2} \erfc z \\
&= \frac{2}{\pi \sqrt{q \gamma}} - \frac{2^{8} \mathcal{N}}{\pi \mathcal{P} \wp_2(d )^4 \gamma^{3/2} \rho^2}  +\mathcal{O}\pr{\frac{1}{\rho^4}}
\label{ClosedFormNodeDegree}
,}
where $z=\sqrt{\frac{ \gamma \mathcal{P} }{\mathcal{N}}} \frac{ \wp_\eta(d )^2 \rho}{16}$ and $\erfc$ is the complementary error function. 
Note that in the limit $\rho\rightarrow \infty$, the mean node degree increases monotonically and approaches the limit $\mu \rightarrow \frac{2}{\pi\sqrt{q\gamma}}$, which is independent of the directivity parameter $d $. 
The second order term seen in the Taylor expansion of \eqref{ClosedFormNodeDegree} however does depend on $d $ and indicates that \textit{the more directional the interfering nodes, the smaller the value of $\wp_\eta(d )$ in the denominator, and hence the lower is the in- and out- mean degree $\mu$.}
This should not come as a surprise as antenna directivity introduces a trade-off between point-to-point link quality (as measured through $H_i$) and point-to-multipoint or multipoint-to-point reach (as measured through $\mu$).
In other words, antenna directivity reduces interference effects since $\mu$, as defined in \eqref{NodeDegreeDenfi}, can be interpreted as a proxy of the signal strength to (or from) a randomly chosen receiver (transmitter); a kind of interference isolation.
Indeed, this is consistent with observations presented in the next section.

%%%%%%%%%%%%%%%%%%%%%%%%%%%%%%
\section{Numerical Simulations \label{sec:num}}

\begin{figure}[t]
\centering
\includegraphics[scale=0.16]{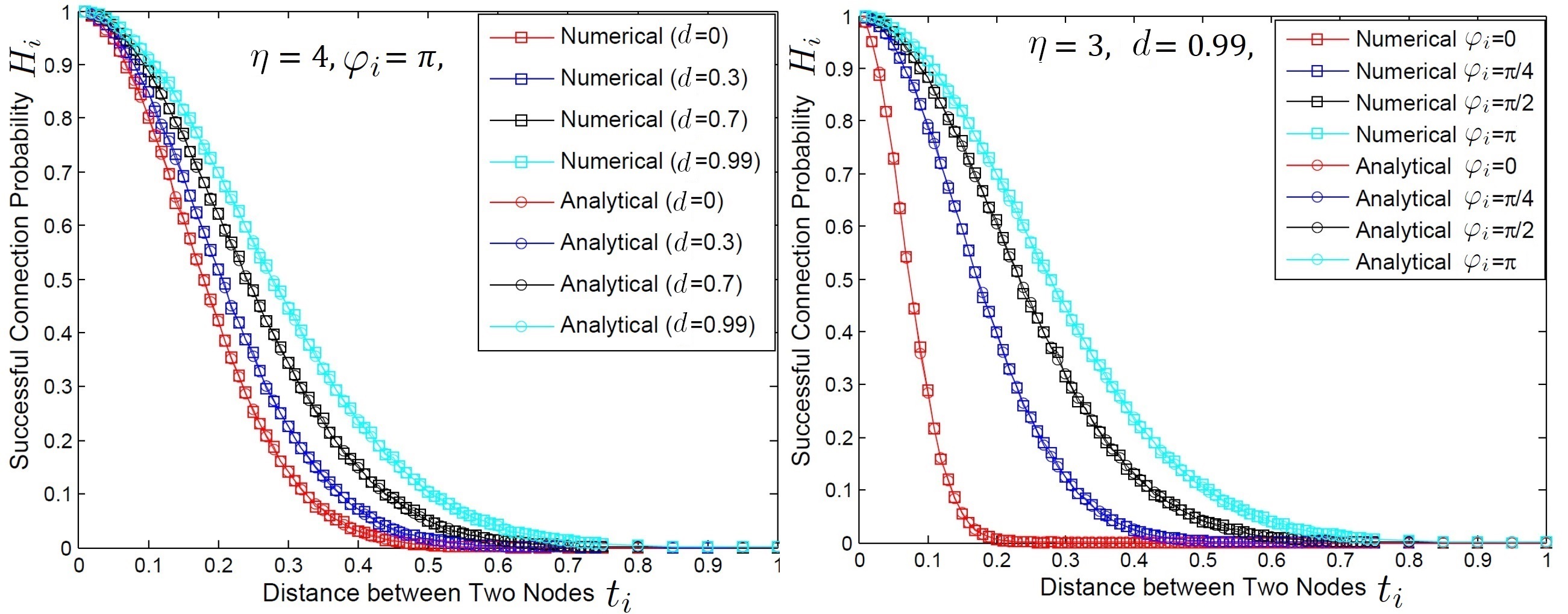}
\caption{
Plots for connection probability $H_{i}$ as a function of the distance $t_i$ between transmitter $i$ and the receiver at the origin.
Computer simulations are compared with the analytic predictions of equation \eqref{ClosedFormHij} for different antenna directivity factors $d $.
\textit{Left:} Transmitter and receiver main radiation lobes are perfectly aligned (i.e. $\varphi_i=\pi$).
\textit{Right:} Transmitter and receiver main radiation lobes are imperfectly aligned (i.e. $\varphi_i\not= \pi$).
}
\label{fig:HijDis}
\end{figure}
\begin{figure}[t]
\centering
\includegraphics[scale=0.164]{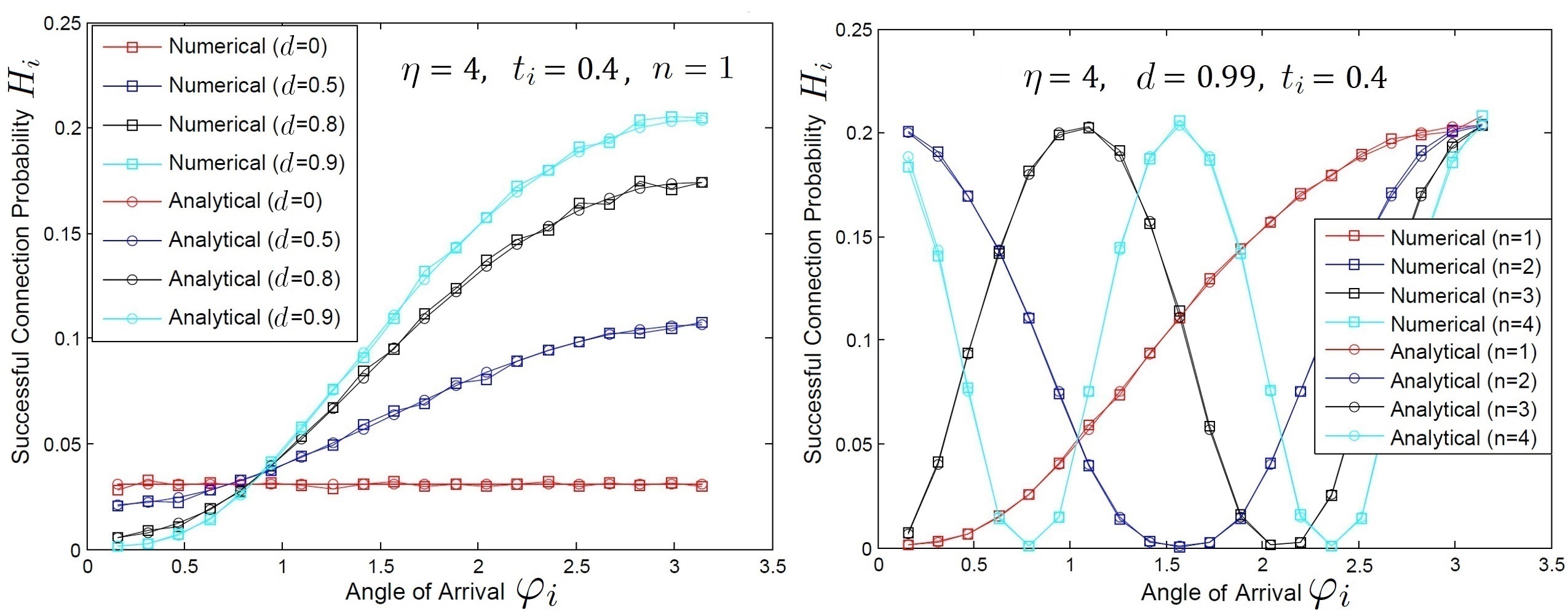}
\caption{
Plots for connection probability $H_i$ as a function of the angle of arrival $\varphi_i$ at a fixed separation distance $t_i = 0.4$. 
Computer simulations are compared with the analytic predictions of equation \eqref{ClosedFormHij} for different antenna directivity factors $d $ (left) and number of antenna radiation lobes $n$ (right).
}
\label{fig:HijTheta}
\end{figure}

In this section, results from computer simulations are presented along with theoretical results from Sec.~\ref{sec:Analysis}. 
Throughout the simulations we set $\mathcal{P} = \mathcal{N} = \rho = q =1, \gamma=0.3,$ and $\epsilon=0$.
To simulate an `infinite domain', a large disk of radius $R=8$ is used and a receiver node is located at the center of the disk. 
%It can be seen that infinite domain is simulated well by using a finite disk area. 
In the simulations, the coordinates of the interfering transmitter are generated uniformly and independently with spatial density $\rho$ inside the disk.
Each transmitter is also assigned a random orientation angle $\varphi_k$ in $[0,2\pi]$.
%Transmitter node is chosen from a certain distance from receiver, which can be seen from x-axis in \ref{fig:HijDis}. 
An additional transmitter is then placed at $(t_i,0)$ with orientation $\varphi_i$.
Channel gains $|h_i|^2$ are randomly generated from an exponential distribution of mean $1$ and the received $\textrm{SINR}_i$ is calculated according to equations \eqref{SINR} and \eqref{Interference}. 
If the $\textrm{SINR}_i$ is larger than a threshold $q$, a successful transmission is achieved.
This process is repeated with different nodes locations, orientations and channel gains for $3\times10^4$ times in a Monte Carlo fashion. 
Therefore, connection probability $H_{i}$ is obtained by dividing number of successful transmissions by the total number transmissions. 
The numerical procedure is repeated for different distances $t_i \in (0,1)$, orientations $\varphi_i$, and directionality factors $d $.

Fig.~\ref{fig:HijDis} compares the numerical simulation results with the analytical expression in \eqref{ClosedFormHij}.
An excellent agreement is observed indicating that a highly directional antenna pattern can help increase the probability of successful connections when transmitter and receiver are aligned. 
In the case where the transmitter-receiver pair are not perfectly aligned a significant degradation in connection quality is observed, especially if the orientation of the a null is towards the receiver. 

Fig.~\ref{fig:HijTheta} compares the dependence of $H_i$ on the orientation $\varphi_i$ of transmitter $i$ at a given fixed distance $t_i=0.4$ but for different directivity factors $d $ and also for the case of multi-lobe antennas with $n=1,2,3,4$. 
Clearly for isotropic antennas $d  =0$, there is no dependence on $\varphi_i$. 
For $d  >0$ however, a significant difference is observed. 
For multi-lobe antenna patterns (e.g., when the transmitter has several active elements or sectorised antennas), the number of peaks experienced in $H_{i}$ as a function of $\varphi_i \in (0,2\pi)$ corresponds to value of $n$ (c.f. Fig.~\ref{fig:direcantenna}).

\begin{figure}[t]
\centering
\includegraphics[scale=0.163]{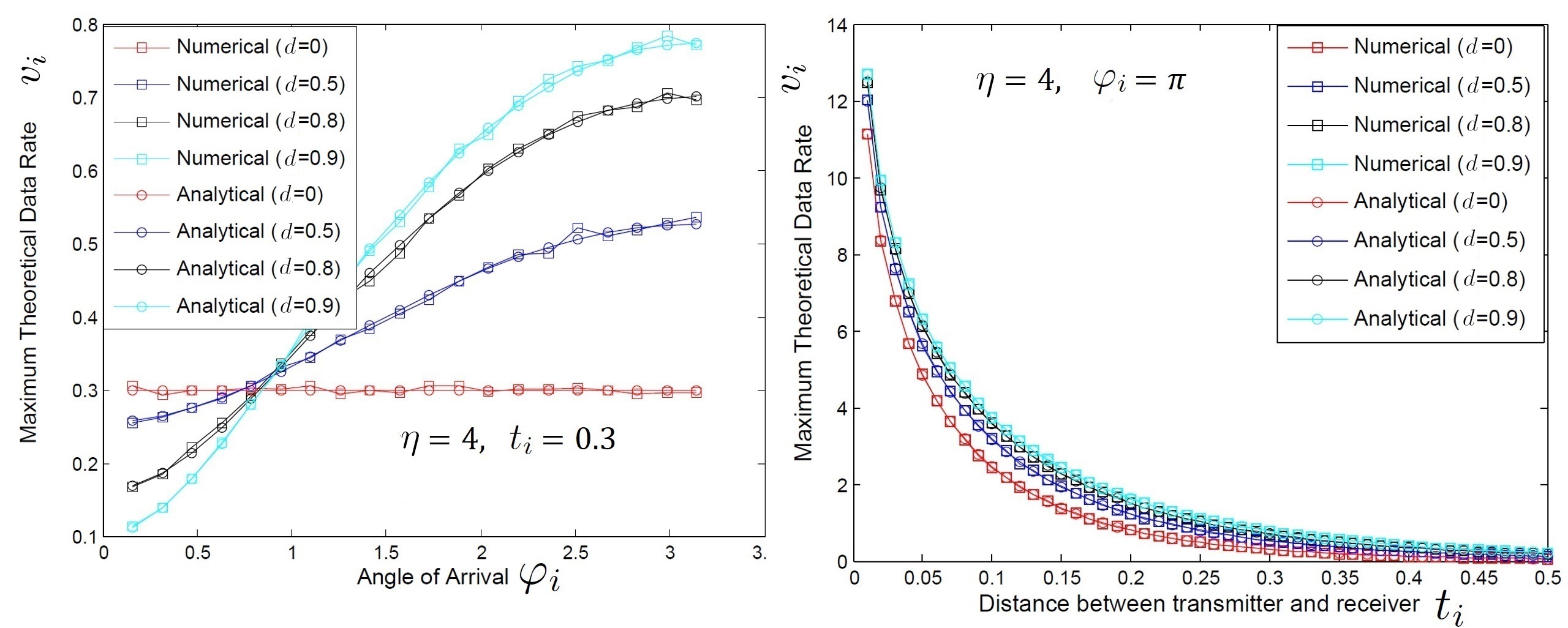}
\caption{
Plots for maximum theoretical data rate $v_{i}$ as a function of arrival angle $\varphi_i$ (left) and separation distance $t_i$ (right).
Numerical results refer to computer simulations as described in the text, whilst the analytical results refer to the numerical integration of \eqref{datarate}.
}
\label{fig:Vij}
\end{figure}

The theoretical maximal data rate is then analysed numerically by applying Shannon capacity equation \eqref{drdefinition} and numerically integrating equation \eqref{datarate}. 
Similar results are obtained as above with excellent agreement as plotted in Fig.~\ref{fig:Vij}.

\begin{figure}
\centering
\includegraphics[scale=0.27]{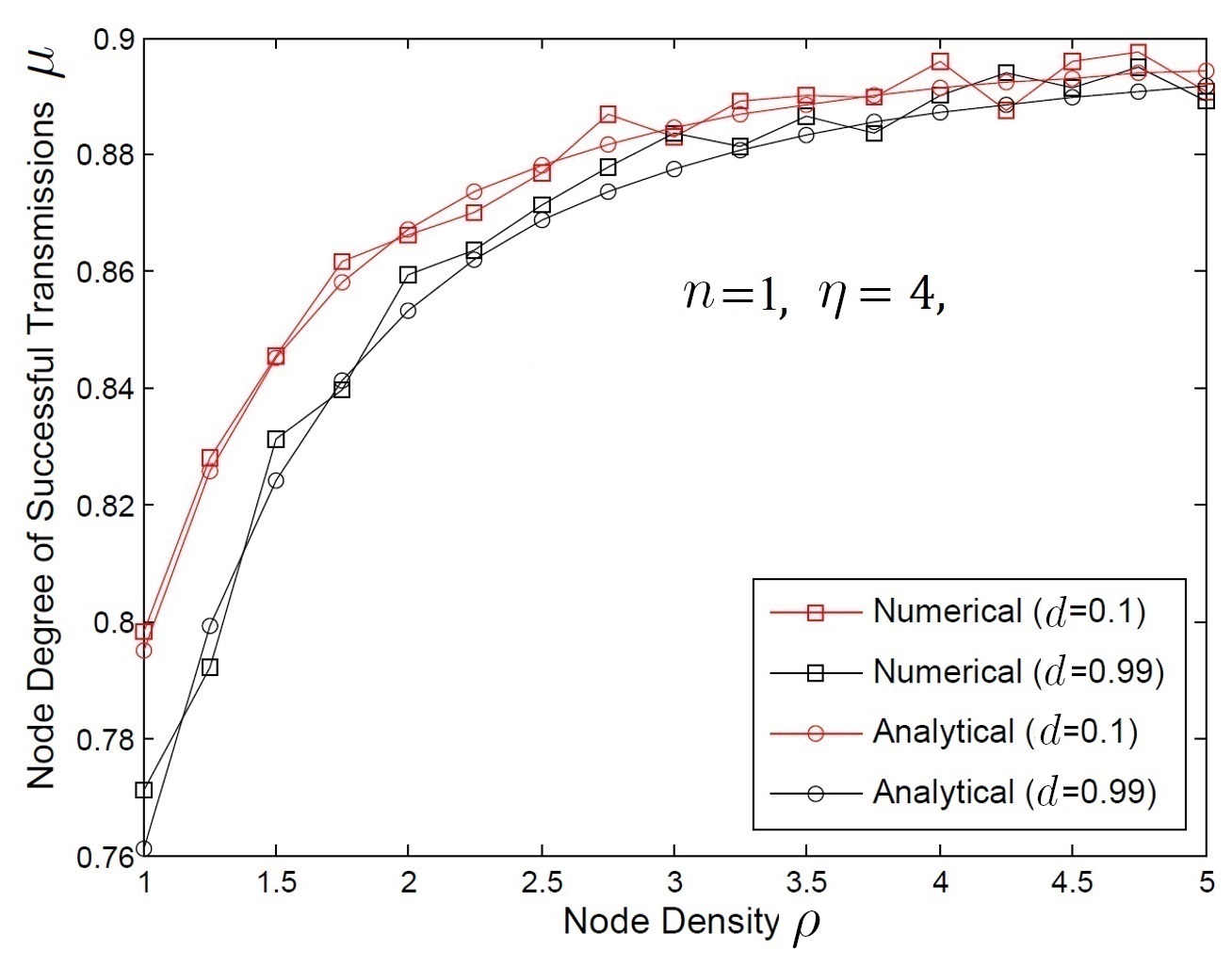}
\caption{
Plots for the mean node degree $\mu$ which measures the expected number of successful transmissions per unit area. 
The analytical curves are obtained using \eqref{ClosedFormNodeDegree}. 
}
\label{fig:muj}
\end{figure}

The mean degree of successful transmissions is also studied numerically.
Here, the transmitter at $(t_i,0)$ is removed, and instead the SINR$_i$ between the receiver at the origin and every other interfering transmitter is calculated for a given realization of locations, orientations, and channel gains.
Each time, the interference term includes all transmitters $k\not=i$.
The algorithm counts the number of transmitters with SINR larger than the threshold $q$ and divides this by the total number of transmitters.
The processes is averaged over $3\times 10^4$ realizations and repeated for different node densities $\rho$ and antenna directionality factors $d $.
The results are plotted in Fig.~\ref{fig:muj} and compared to the analytical predictions of the mean degree given by equation \eqref{ClosedFormNodeDegree}.
An excellent agreement is observed confirming our analysis and conclusions.

%%%%%%%%%%%%%%%%%%%%%%%%%%%%%%
\section{Discussion and Conclusions
\label{sec:con}}

In the absence of interference, randomised beamforming schemes (or directional antennas with random orientations) can improve the connectivity of ad hoc networks by reducing the number of hops needed for any two nodes to communicate in a multi-hop fashion \cite{bettstetter2005does,georgiou2015multihop}.
In this paper we allow for interference caused by concurrent transmissions and investigate local network observables such as outage probability, connectivity, capacity, and mean node degree for which we derive novel analytic expressions in closed-form using tools from stochastic geometry \cite{haenggi2009stochastic}.
It is shown that ad hoc networks equipped with directional antennas benefit from interference isolation on both the receiving and transmitting ends. 
The former refers to a receiver whose antenna gain is oriented towards a transmitter and away from potential interferes, whilst the latter refers to interfering transmitters whose antenna gains ``miss" the receiver.
This physical picture is verified through computer simulations in Sec. \ref{sec:num} but also quantified by the analytic results presented in Sec. \ref{sec:Analysis}.

We have strengthened the case for the use of directional antennas in ad hoc networks and provided closed-form expressions enabling wireless communication parameters to be tuned more appropriately to meet network design criteria.
It would be interesting to see experimental verification of our predictions in a random access scheme where no \textit{a priori} node location information is available or in random uncontrolled (e.g. air-dropped) WSN deployments.

%%%%%%%%%%%%%%%%%%%%%%%%
\section*{Acknowledgements}

The DIWINE consortium would like to acknowledge the support of the European Commission partly funding the DIWINE project under Grant Agreement CNET-ICT-318177. 
The authors would like to thank the Directors of the Toshiba Telecommunications Research Laboratory for their support.

%%%%%%%%%%%%%%%%%%%%%%

\bibliographystyle{ieeetr}
\bibliography{mybib}

\end{document}